\documentstyle[aasms4, psfig]{article}

\newcommand\psr{\hbox{PSR~J0631+1036}}
\newcommand\Einstein{{\it Einstein}}
\newcommand\einstein{{\it Einstein}}
\newcommand\ASCA{{\it ASCA}}
\newcommand\asca{{\it ASCA}}
\newcommand\ROSAT{{\it ROSAT}}
\newcommand\rosat{{\it ROSAT}}
  
\received{}
\accepted{}
\journalid{}{}
\articleid{}{}

\lefthead{Torii \etal}
\righthead{{\it ASCA} Detection of Pulsed X-ray Emission from PSR J0631+1036}

\begin{document}

\title{{\it ASCA} Detection of Pulsed X-ray Emission from PSR~J0631+1036}

\author{Ken'ichi Torii$^1$, Y. Saito$^2$, F. Nagase$^2$,
T. Yamagami$^2$,\\ T. Kamae$^3$, M. Hirayama$^4$, N. Kawai$^{1, 5, 6}$,
I. Sakurai$^{1}$, M. Namiki$^1$, 
S. Shibata$^7$, S. Gunji$^7$,\\ and J.P. Finley$^8$}

\altaffiltext{1}{Cosmic Radiation Laboratory, Institute of
Physical and Chemical Research (RIKEN), 2-1, Hirosawa, Wako,
Saitama, 351-0198, Japan}
\altaffiltext{2}{Institute of Space and Astronautical Science, 3-1-1 Yoshinodai, Sagamihara, Kanagawa, 229-8510, Japan; saito@balloon.isas.ac.jp}
\altaffiltext{3}{Graduate School of Science, Hiroshima University,
1-3-1 Kagamiyama, Higashihiroshima, 739-8526, Japan; torii@crab.riken.go.jp
}
\altaffiltext{4}{Santa Cruz Institute for Particle Physics, University 
of California at Santa Cruz, Santa Cruz, CA 95064, USA
}
\altaffiltext{5}{
Space Utilization Research Program (SURP), Tsukuba Space Center (TKSC), National Space Development Agency of Japan (NASDA), 2-1-1 Sengen, Tsukuba, Ibaraki, 305-8505, Japan
}
\altaffiltext{6}{Department of Physics, Tokyo Institute of Technology,
2-12-1 Ookayama, Meguro-ku, Tokyo, 152-8551
}
\altaffiltext{7}{Department of Physics, Yamagata University,
1-4-12 Kojirakawa, Yamagata, 990-8560, Japan
}
\altaffiltext{8}{Department of Physics, Purdue University, West
Lafayette, IN 47907, USA
}

\begin{abstract}

{\it ASCA}'s long look at the 288~millisecond radio pulsar,
PSR~J0631+1036, reveals coherent X-ray pulsation from this source for
the first time. The source was first detected in the serendipitous
{\it Einstein} observation and later identified as a radio
pulsar. Possible pulsation in the gamma-ray band has been detected
from the {\it CGRO} EGRET data (Zepka, et al. 1996).  The X-ray
spectrum in the \asca\ band is characterized by a hard power-law type
emission with a photon index $\simeq 2.3$, when fitted with a single
power-law function modified with absorption. An additional blackbody
component of $kT \simeq 0.14$~keV increases the quality of the
spectral fit. The observed X-ray flux is $2.1\times 10^{-13}\, {\rm
ergs\, s^{-1}\, cm^{-2}}$ in the 1-10~keV band. We find that many
characteristics of PSR~J0631+1036 are similar to those of middle-aged
gamma-ray pulsars such as PSR~B1055-52, PSR~B0633+17 (Geminga), and
PSR~B0656+14.

\end{abstract}

\keywords{pulsars: general --- pulsars: individual
(PSR~J0631+1036) --- X-rays:  stars: neutron}

\section{Introduction}

Studies of rotation powered pulsars in the X-ray band give us
information of their evolution of magnetospheric activities, surface
temperatures, and interaction of the pulsar wind with the surrounding
medium (e.g., Becker \& Tr${\rm {\ddot u}}$mper 1997; Saito 1998). The
most energetic Crab-like pulsars with the spin-down power of ${\dot E}
\stackrel{>}{_\sim} 10^{37}\, {\rm ergs\,s^{-1}}$ show X-ray pulsations
of magnetospheric origin. Most of them are associated with a supernova
remnant and power synchrotron nebulae as a result of shock interaction
of the pulsar wind with the ambient medium. The Vela-like pulsars,
characterized by their spin-down power of
$10^{36} {\rm
ergs\,s^{-1}} \stackrel{<}{_\sim}{\dot E} \stackrel{<}{_\sim}  10^{37} 
{\rm ergs\,s^{-1}}$, are embedded in the
extended synchrotron nebulae which often makes it difficult to study
the neutron star itself. Older and weaker
($10^{34} {\rm
ergs\,s^{-1}} \stackrel{<}{_\sim}{\dot E} \stackrel{<}{_\sim}10^{35} {\rm
ergs\,s^{-1}} $) sources show pulsating
blackbody-type spectra below $\simeq 0.5-0.6$~keV, as well as the
pulsating power-law type spectra in the hard energy band. Typical
objects of this class are the three Musketeers, PSR~B1055-52,
PSR~B0656+14, and PSR~B0633+17 (Geminga) (e.g., Finley, ${\rm {\ddot
O}}$gelman, \& Kizilo\v{g}lu 1992; Greiveldinger, et al. 1996; Becker, et al. 1999). Interestingly,
these middle-aged ($\tau \simeq 10^5-10^6$~yrs) pulsars convert their
spin-down power into high energy $\gamma$-ray photons with
higher efficiency than in younger pulsars (e.g.,
Thompson 1996; Kifune 1996)

 The source studied herein, \psr, was discovered by the targeted
searches for radio pulsars in unidentified 
\einstein\ X-ray sources (Zepka, et al. 1996).
The pulse period, $P\simeq 0.288$~s, and its derivative, ${\dot
P}\simeq 1.0\times 10^{-13}\, {\rm s\, s^{-1}}$, give the 
characteristic age of ${\tau}=P/(2{\dot P})=4.3\times 10^4$~yrs. The
spin-down power is ${\dot E}=5.4\times 10^{34}\, I_{45}\, {\rm ergs\,
s^{-1}}$ where $I_{45}$ is the moment of inertia of the neutron star
normalized to $10^{45}\, {\rm g\, cm^2}$. This range of parameters is
between the Vela-like pulsars and the three Musketeers.

In the \Einstein\ IPC data, about 50 photons were detected from the
X-ray counterpart of the radio pulsar \psr. The spectral fit gave a
blackbody temperature of $kT=0.27\pm 0.08$~keV, an absorption column
$N_H=(9\pm 4) \times 10^{21}\, {\rm cm^{-2}}$, and an observed X-ray
flux $F_X = 1.9\times 10^{-13}\, {\rm ergs\,s^{-1}\, cm^{-2}}$ in the
$0.16-3.5$~keV band (Zepka, et al. 1996). Unfortunately, the source
was located on the support rib of the IPC field. This made the source
position and source spectra rather uncertain. Later, the source was
serendipitously observed in the field of view of \ROSAT\ PSPC. Spectral
analysis gave a best-fit blackbody temperature $kT=0.18\pm 0.08$~keV,
an absorption column $N_H = (1.2 \pm 0.6) \times 10^{21}\, {\rm
cm^{-2}}$, and an effective radius of the emitting region of
approximately $R_{\rm BB}\simeq 1$~km (Zepka, et al. 1996). Again, the
source was unfortunately shadowed by the detector supporting ribs,
making the obtained spectral parameters rather uncertain.

Interestingly, it was found that this source might be pulsating in the
$\gamma$-ray energy band. Zepka, et al. (1996) folded the arrival times
of 267 counts from {\it CGRO} EGRET at the
expected pulse period. They found that the folded
light curve was significantly displaced from uniform distribution at
more than 99~\% confidence.

\section{Observation}

  We have proposed and carried out the \asca\ observation of \psr\
during 1998, October 16-18.
\ASCA\ (Tanaka,
Inoue, \& Holt, 1994) carries two kinds of X-ray detectors at the
foci of four identical X-ray telescopes. The X-ray CCD camera, SIS
(Solid-state Imaging Spectrometer; Burke, et al. 1994), has the higher
energy resolution and relatively high detection efficiency in the soft
energy band. The time resolution of SIS is 4-16~s in the standard mode
which is not suitable for timing observation of fast pulsars.  The
imaging gas-scintillation proportional counter, GIS (Gas Imaging
Spectrometer; Ohashi, et al. 1996; Makishima, et al. 1996), has
relatively high detection efficiency in the hard energy band and high time
resolution.
We operated the SIS in 1-CCD faint mode with 4-s time
resolution. Therefore, only pulse phase averaged spectroscopy could be
made with the SIS. We operated the GIS in the PH mode and assigned a
part of the telemetry bit to increase the time resolution reducing the
spectral information. The resultant time resolution was 3.9~ms or
better depending on the telemetry rate. We used screened event data
according to the standard REV2 processing (Pier 1997). The effective exposure time
was 69.6 and 76.1~ks for each SIS and GIS, respectively, and the net
time span of observation was $T=160$~ks.

\section{Analysis}

\subsection{Timing}

We analyzed the GIS data to search for pulsations in the X-ray
band. The GIS observation began at 51103.0185~MJD and ended at
51104.8260~MJD.  The data from GIS2 and GIS3 were co-added, yielding
$\simeq 1100$ events within 3$'$ radius circular region in the total
energy band of 0.7-7~keV, including background. The photon arrival
times were barycentered and $z_1^2$ test (Buccheri, et al. 1983) and
epoch folding search (e.g., Leahy, et al. 1983) were applied
bracketing the expected period, 
 $P_{\rm exp} = 0.2877671$~s
at
MJD~51103.9305 (middle time of the current \asca\ observation at
barycenter), from the radio ephemeris effective during the current
observation (Nice 2000, private communication).

Figure 1a shows the result of $z_1^2$ test. We can clearly see a peak
at the expected pulse period.  No similar peaks have been found in the
background data with better statistics extracted from the same observation. 
We then applied the epoch folding search. The events were folded into
6~bins and $\chi^2$ values were calculated from each trial period. A
significant peak is found (Figure 1b) at the period consistent with
that obtained from the $z_1^2$ test.  The probabilities to find higher
peaks out of random fluctuations with a single trial are as low as
$1.8\times 10^{-6}$ for $z_1^2$ test and $1.1\times 10^{-6}$ for
folding search. Considering the reasonable number of independent
trials ($\simeq 18$) as
discussed below, the chance probabilities to find higher peaks 
are $3.2\times 10^{-5}$ for the $z_1^2$ test and $2.0\times
10^{-5}$ for the folding search.
Therefore, we conclude that we have significantly detected X-ray pulsations
from \psr\ for the first time. 

The pulse period is determined from folding
search to be $P=0.2877672 (1)$~s at MJD~51103.9305. Here, the value
in parenthesis is a 90~\% confidence error to the last digit. We
estimated this error by using the method of Leahy (1987), which gave
tighter constraint than the nominal error of $P^2/(2T)=3\times
10^{-7}$~s.

The detected pulse period is consistent with the effective radio
ephemeris while it is significantly shorter than the period, $P =
0.2877676$~s, which is extrapolated from the previous radio ephemeris
(Zepka, et al. 1996) by $\Delta P=-4\times 10^{-7}$~s. This difference
in period may not be ascribed to a finite value of negative ${\ddot
P}$, since the corresponding value of ${\ddot P}\simeq -3\times
10^{-23}\, {\rm s^{-1}}$ gives an unreasonably large braking index,
$n\simeq 7\times 10^2$. Although it is not known if \psr\ has recently
glitched or not due to lack of published results (e.g., Johnston \&
Galloway 1999), we consider that the value of 
$\Delta P/P = -1.4\times 10^{-6}$ is
naturally interpreted as the result of glitch activity.
The reasonable number of independent trials for the current timing analysis is
then estimated by considering a large
glitch of $|\Delta P/P| \stackrel{<}{_\sim} 10^{-5}$ during the
4.3~years between the last published radio observation (Zepka, et al. 1996) and
the \asca\ observation.  The largest glitch
ever observed would fall within the range considered here (Wang, et
al. 2000). Then the number of independent Fourier bins covering the
period range, $P_{\rm exp}-|\Delta P_{\rm exp}|<P<P_{\rm
exp}+|\Delta P_{\rm exp}|$, is 18.

The top panel of figure 2 shows the folded light curves in the total energy
band. The pulse shape is singly peaked and sinusoidal. The pulse
amplitude is $45\pm 16$~\% of the source flux excluding background.
Comparison
has been made of pulse profiles in different energy bands. 
The pulse amplitudes, derived by fitting the profiles with a sinusoidal
curve, are $31\stackrel{+12}{_{-17}}$~\% and $63\pm 18$~\% in 0.7-1.9
and 1.9-7~keV band, respectively.

\subsection{Spectrum}

We analyzed the pulse phase averaged spectra by using both the SIS and GIS
data. Spectra were extracted from circular regions of $3'$
radius. Background was subtracted from the same observation. For the
SIS, events from the whole chip excluding the circular region of $4'$
radius around the source were used as background. For the GIS, events
from circular regions of $7.5'$ radius to the north-east of the 
source were used as background.

A single power-law function with soft X-ray absorption was first
applied for spectrum from each detector. This simple model gave
acceptable fit to each data set and the spectral parameters were found
to be consistent with each other within statistical errors. Therefore,
we simultaneously fit the two SIS and two GIS spectra. In
the fitting procedure, each data point was weighted according to the
statistics (number of photons in each bin) taking into account the 
propagation of error due to background subtraction.
For clarity,
figure 3 shows
the representative spectra and the best-fit model function for the 
two detectors (SIS1 and GIS3).
The
best-fit parameters are the power-law photon index, $\gamma =
2.3\stackrel{+0.5}{_{-0.4}}$, the absorbing hydrogen column density, $N_H =
(0.2\stackrel{+0.2}{_{-0.1}})\times 10^{22}\,\, {\rm cm^{-2}}$, and
the normalization, $7\stackrel{+3}{_{-2}} \times 10^{-5}\,\, {\rm
photons\,\, keV^{-1}\,\, s^{-1}\,\, cm^{-2}}$ at 1~keV with
$\chi^2$/dof=140.6/112. These errors are at one parameter 90\%
confidence.  The observed flux is $f_X = 1.9\times 10^{-13}\, {\rm
ergs\, s^{-1}\, cm^{-2}}$ and the intrinsic flux is $f_X = 2.0\times
10^{-13}\, {\rm ergs\, s^{-1}\, cm^{-2}}$, corresponding to the
luminosity of $L_X = 2.4\times 10^{31}\, d_1^2 \,\Omega_{4\pi}\, {\rm
ergs\, s^{-1}}$, all in 1-10~keV range. Here, $d_1$ is the source distance
normalized to 1~kpc and $\Omega_{4\pi}$ is the emitting solid angle
normalized to $4\pi$ steradian.  A single blackbody model with
absorption gave the best-fit temperature of $kT=0.43$~keV while the
fit was not acceptable with $\chi^2$/dof=178.9/112.

Since the soft X-ray emission was well fit by a blackbody model
(Zepka, et al. 1996), we tried blackbody plus power-law model, modified
by a single absorption component. When the blackbody
temperature and the normalization were fixed at the best-fit parameters
derived from \rosat\ PSPC, $kT=0.18$~keV and $R_{\rm BB}=1$~km at
1~kpc, the following parameters are obtained. The power-law photon
index, $\gamma= 1.2\pm 0.4$, the normalization for the power-law
component $Norm=2.1\stackrel{+1.4}{_{-0.9}} \times 10^{-5}\, {\rm photons\,
keV^{-1}\, s^{-1}\, cm^{-2}}$ at 1~keV and the absorption column $N_H
= 0.96\stackrel{+0.08}{_{-0.07}}\times 10^{22}\, {\rm cm^{-2}}$. In
this case, crossover of the two components occurs at around
1.9~keV. The quality of the fit is slightly worse than the single
power-law model, $\chi^2/dof =  147.1/112$. When the emitting radius of 
blackbody is set free, $R_{\rm BB}=(0.6\stackrel{+0.1}{_{-0.2}})\, d_1$~km is
obtained with $\gamma= 1.7\pm 0.4$, $Norm=(4\stackrel{+2}{_{-1}})\times 10^{-5}\, {\rm photons\,
keV^{-1}\, s^{-1}\, cm^{-2}}$, $N_H
= 0.5\pm 0.2\times 10^{22}\, {\rm cm^{-2}}$, and $\chi^2/dof =  128.2/111$.

Starting with the best-fit parameters for the blackbody component
obtained from \rosat, we tried to set all the five parameters free in
the two component (blackbody plus power-law) model. However, the
parameters could not be well constrained due to trade-offs between the
two continuum components and the arbitrary absorption
column. Therefore, we fixed the temperature at a value in the range
$0.1\, {\rm keV} \leq kT \leq 0.3\, {\rm keV}$ with every 0.01~keV
step. Then the minimum of $\chi^2/dof = 126.1/111$, was obtained with
$kT= 0.14$~keV and $R_{\rm BB}=(2.1\stackrel{+0.5}{_{-0.6}})\,
d_1$~km. The $\chi^2$ value with $kT=0.14$~keV is thus smallest and we
consider this model is most suitable for the current \asca\ data. The
other spectral parameters for $kT=0.14$~keV are the power-law photon
index, $\gamma= 1.9 \pm 0.4$, the normalization for the power-law
component $6 \stackrel{+3}{_{-2}} \times 10^{-5}\, {\rm photons\,
keV^{-1}\, s^{-1}\, cm^{-2}}$ at 1~keV and the absorption column $N_H
= (0.8 \pm 0.2) \times 10^{22}\, {\rm cm^{-2}}$. With this model, the
total observed X-ray flux is $f_X = 2.1\times 10^{-13}\, {\rm ergs\,
s^{-1}\, cm^{-2}}$ and the intrinsic flux is $f_X = 3.5\times
10^{-13}\, {\rm ergs\, s^{-1}\, cm^{-2}}$ in the 1-10~keV band. The
corresponding luminosity is $4.2\times 10^{31}\, d_1^2 \,
\Omega_{4\pi}\, {\rm ergs\, s^{-1}}$. The blackbody component
contributes 16~\% and 34~\% of the observed and intrinsic flux,
respectively.

 We then examined the pulse phase dependence of spectral shape by
using the GIS data. We divided the data into two subsets, pulse on
(high intensity) phase and pulse off (low intensity) phase. The two
phases are split at phase 0.0625 and 0.5625 in figure 2.  Since the
statistics are limited, we fixed the absorption column at the best-fit
value obtained from a single power-law model and fit the spectra with
a single power-law function. The photon indices were $\gamma =
2.5\stackrel{+0.7}{_{-0.5}}$ and $\gamma =
3.5\stackrel{+1.0}{_{-0.9}}$ for pulse on and off phases,
respectively. Although harder spectrum is suggested for
higher flux phase, the photon indices are consistent within
statistical errors.

\section{Discussion}

 The overall properties of the multi-wavelength spectrum of \psr\ from
radio to X-ray and $\gamma$-ray are similar to those of other
$\gamma$-ray pulsars (Thompson 1996). They are radiating a large
fraction ($\stackrel{>}{_{\sim}}0.1$) of their spin-down power in the
$\gamma$-ray band. The X-ray luminosity, $L_X=4.2\times 10^{31}\,
d_1^2 \,\Omega_{4\pi}\, {\rm ergs\, s^{-1}}$, obtained from the two
component spectral fit, is $\simeq 0.08\, d_1^2\, \Omega_{4\pi}\,
I_{45}^{-1}\, \%$ of the spin-down power, which is comparable to those
of Vela pulsar and the three Musketeers (e.g., Saito 1998).  For
\psr\, the power-law photon index which smoothly connects the X-ray
and $\gamma$-ray flux is $\gamma \simeq 1.2$. Therefore, the simple
extension of the X-ray spectrum with the photon index, $\gamma=1.9\pm
0.4$, obtained from the best power-law plus blackbody model, predicts
lower $\gamma$-ray flux than observed. This situation may be
reconciled with complete pulse phase spectroscopy including
appropriate blackbody model, which should be done in future
observations.

The observed pulse fraction, 45~\% in the 0.7-7~keV band, is
comparable with $\simeq 40$~\% found for PSR~B1055-52 above 1~keV (${\rm
{\ddot O}}$gelman \& Finley 1993; Figure 1b) and $\simeq 55$~\% found
for Geminga in 1-4~keV band (Halpern \& Wang 1997). While these values
are significantly larger than the pulse fraction of $14\pm 2$~\% for
PSR~B0656+14 as measured by \rosat\ (Finley, et al. 1992). This might
be partly due to the energy dependent pulse fraction of
PSR~B0656+14.

The significant difference between the detected and extrapolated pulse
period may be understood as a result of glitch activity. 
Urama \& Okeke (1999) have found that there exists a good correlation
between young pulsars' spin-down rate and glitch activity.  They predict the
interval between Vela-size glitches of average $\Delta P/P=-2\times
10^{-6}$ to be 7~years for \psr. In terms of the glitch activity
parameter $A_g$, the mean fractional change in period per year owing
to glitches, the predicted value is $A_g=3.09\times 10^{-7}\, {\rm
yr^{-1}}$ (Urama \& Okeke 1999). This corresponds to the accumulated
period change during the 4.3~yrs of $\Sigma (\Delta P / P) \simeq -
1.3\times 10^{-6}$. These predictions are in good agreement with those
observed, $A_g = 3.2\times 10^{-7}\, {\rm yr^{-1}}$ or $\Sigma (\Delta
P / P) = - 1.4\times 10^{-6}$.

 On the other hand, Lyne, Shemar, \& Graham-Smith (2000) studied the
statistical properties of pulsar glitches by using well-defined sample
and found a strong indication that pulsars with large magnetic fields
suffer many small glitches while others show a smaller number of large
glitches. Comparing the period and its derivative for \psr\ with those
in their sample, PSR~B1758-23, possibly associated with the supernova
remnant W28 (Kaspi, et al. 1993), has similar range of
parameters. Therefore, \psr\ might have experienced a large number of
medium size ($\Delta P/P\simeq -10^{-7}$) glitches as in the case of
PSR~B1758-23. Lyne et al. (2000) also found a good correlation between
${\dot \nu}$ and the glitch spin-up rate, ${\dot \nu_{\rm glitch}}$,
defined as the cumulative effect of glitch upon the frequency
derivative. The relation, ${\dot \nu_{\rm glitch}} = -0.017 {\dot \nu}$,
leads to $\Delta P = -2.4\times 10^{-7}$~s for \psr\ during the time
span of 4.3~yrs, in reasonable agreement with that observed, $\Delta
P=-4\times 10^{-7}$~s.

PSR~J0631+1036 was first detected in the soft X-ray band, since it is
in the direction of the Galactic anti-center ($l$,
$b$)=($201^\circ.22$, $0^\circ.45$) where the interstellar absorption
is much smaller than in the Galactic plane toward the Galactic center. 
Also, the detection of pulsed X-ray emission could be made only with
long exposure time as performed with \asca\ referring to the
previously known radio period. This lesson suggests that there may be
other similar sources hidden in the Galactic plane.  Combined analyses
of radio, X-ray, and unidentified $\gamma$-ray sources may be useful
(e.g., Roberts, Romani, \& Kawai 2001).

In summary, we have detected the pulsed X-ray emission from \psr\ for
the first time. The negative offset of the observed period from the
extrapolated radio ephemeris is attributable to the
accumulated change in period due to glitch activity. The X-ray
spectrum is well described by a power-law plus blackbody model with
observed flux $f_X = 2.1\times 10^{-13}\, {\rm ergs\, s^{-1}\,
cm^{-2}}$ in the 1-10~keV band.

\begin{acknowledgements}
{\noindent Acknowledgments} --- 
The authors would thank David Nice for providing us with the radio
ephemeris of PSR J0631+1036 prior to publication.
The authors are grateful to all the members of the {\it ASCA} team for 
making the observation possible.

\end{acknowledgements}

\clearpage

\begin{figure}
\centerline{
\psfig{figure=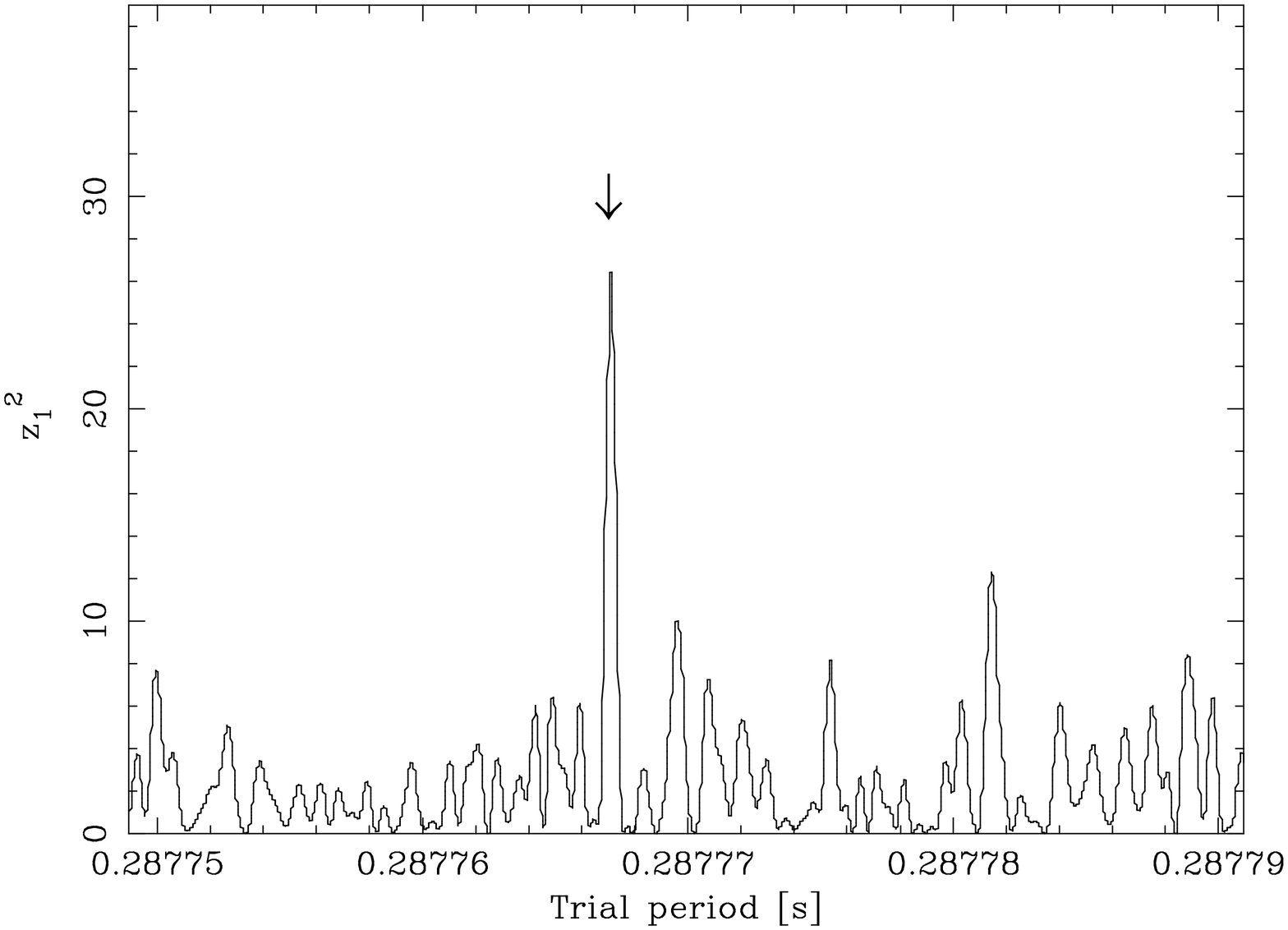,height=7.0cm,angle=270} 
\psfig{figure=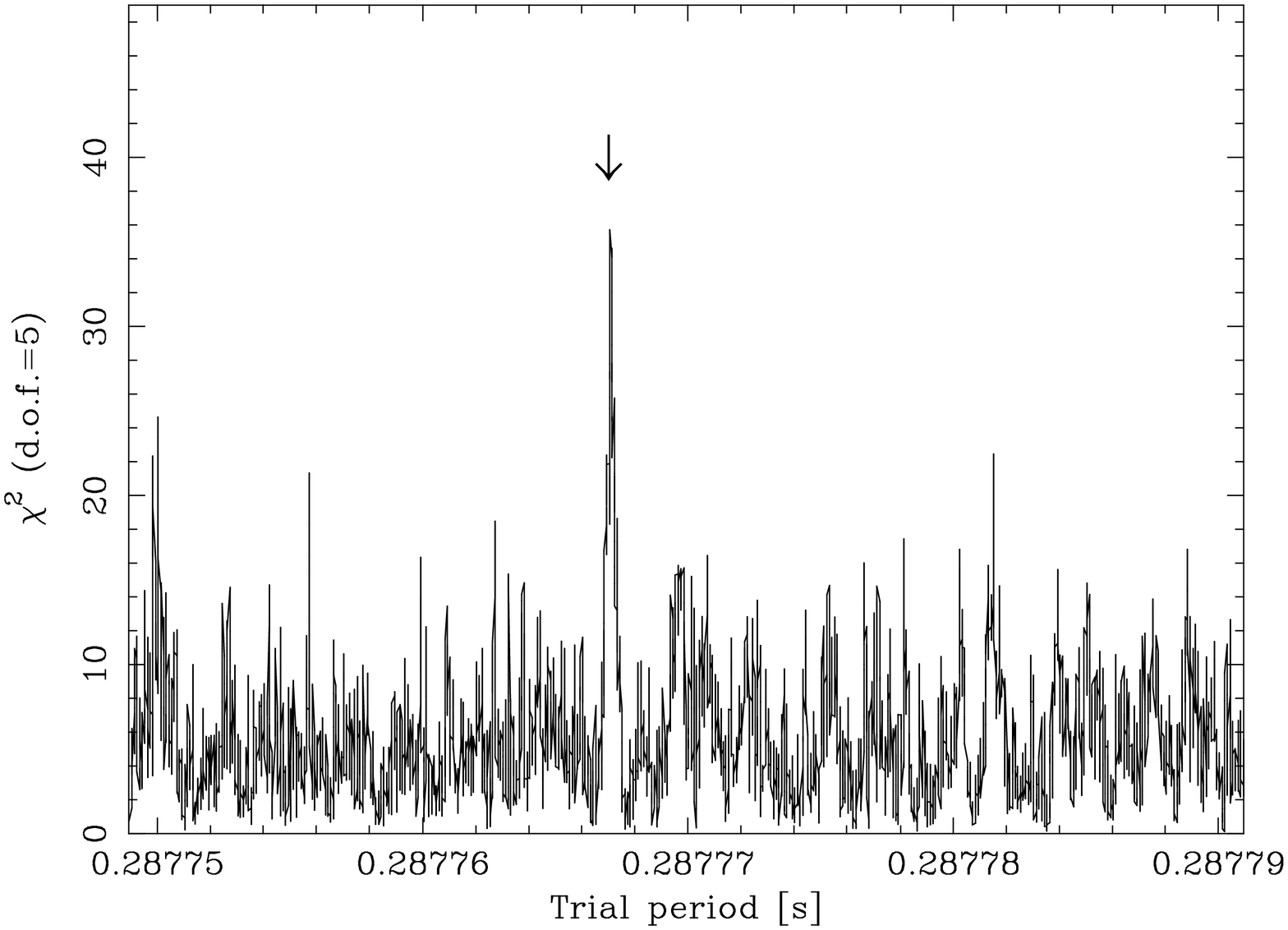,height=7.0cm,angle=270} 
}
\bigskip
\caption{The $z_1^2$ and $\chi ^2$ are plotted as a function of trial
periods. Arrows show the expected period from the effective radio ephemeris. }
\end{figure}

\begin{figure}
\centerline{
\psfig{figure=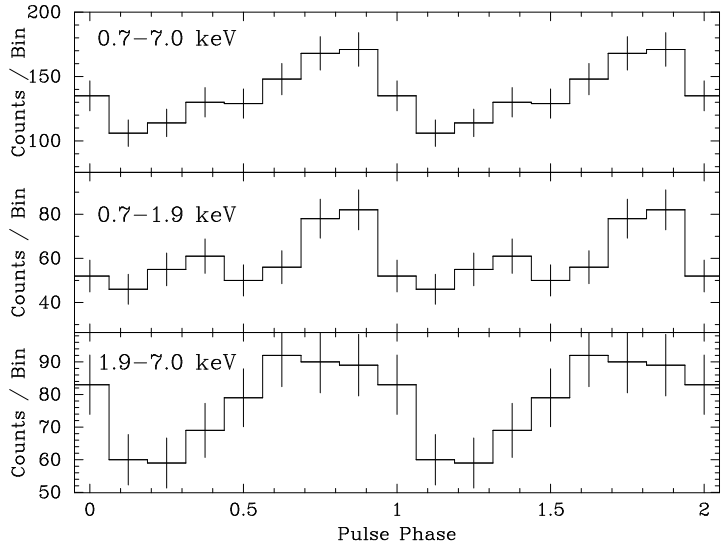,height=8.0cm,angle=270} 
}
\bigskip
\caption{The folded pulse profiles are shown for the full energy range 
0.7-7.0 ({\it top panel}) and two sub ranges 0.7-1.9 ({\it middle panel}) and 1.9-7.0 keV
({\it bottom panel}). The vertical axes show the number of events in
each bin, and they are displayed so that the background levels 
correspond to the baseline of each panel. Two complete cycles are plotted for
clarity.}
\end{figure}

\begin{figure}
\centerline{
\psfig{figure=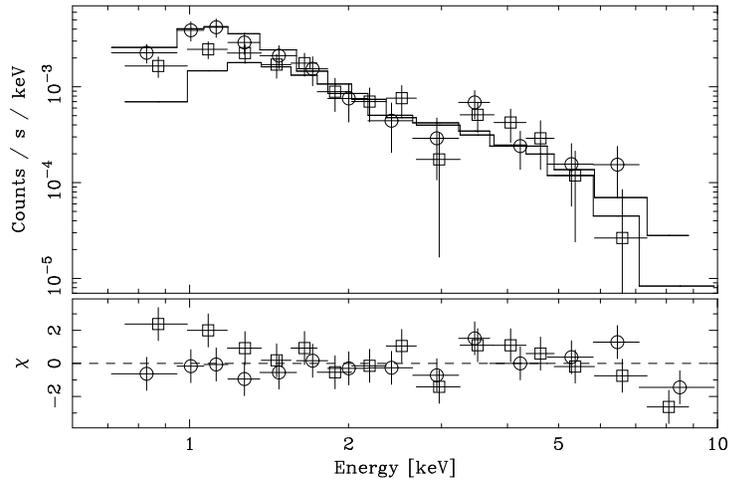,height=8.0cm,angle=270}
}
\bigskip
\caption{{\it Top panel}:  The observed energy spectra and best-fit absorbed power-law
function are shown for the two representative detectors. {\it Bottom panel}: 
Residuals normalized by the standard deviations are
shown. Circles and squares show SIS1 and GIS3 data, respectively.}
\end{figure}

\end{document}